\documentclass{llncs}
\usepackage{graphicx}
\usepackage{eurosym}

\begin{document}

\title{Wextractor: Follow-up of the evolution of prices in web pages}
%%\thanks{}}

\author{Jorge Lloret-Gazo}
\authorrunning{Jorge Lloret-Gazo}   % abbreviated author list (for running head)
%
%%%% modified list of authors for the TOC (add the affiliations)
\tocauthor{Jorge
Lloret-Gazo(Universidad de Zaragoza)}

\institute{Dpto. de Inform\'atica e Ingenier\'{\i}a de Sistemas.\\
Facultad de Ciencias. Edificio de Matem\'aticas.\\ Universidad de
Zaragoza. 50009 Zaragoza. Spain.\\ \email{jlloret@unizar.es}}
\maketitle   
\begin{abstract}
In the e-commerce world, the follow-up of prices in detail web pages
is of great interest for things like buying a product when it falls below
some threshold. For doing this task, instead of bookmarking the pages
and revisiting them, in this paper we propose a novel
web data extraction system, called Wextractor. It consists
of an extraction method and a web app for listing the retrieved prices.
As for the final user, the main feature of Wextractor is usability
because (s)he only has to signal the pages of interest and
our system automatically extracts the price from the page.
\end{abstract}

\section{Introduction}

Web users are interested in finding the price
of an entity in different web sites and in following it up
during a certain period of time for different purposes. 
For example, an user is interested in the price of a pair of shoes in 
different web sites with the purpose of buying them  when
they fall below some threshold due, for example, to discounts
or special offers.

A manual solution to this problem is to bookmark
the pages of interest and to revisit them
again and again until the desired price is found.
Instead of this boring and repetitive task, 
we propose to use our web data extraction sytem,
called Wextractor, which consists of an extraction method and a web app
for listing the retrieved prices. The extraction method
takes as input the urls of the pages of interest and, upon
user request, it retrieves again the price
in all the pages of interest.

Our proposal for extraction fits into the automatic extraction 
technique as described in~\cite{DBLP:books/sp/dcsa/Liu07}.
It is composed of two kinds of extraction methods. The first is called {\em from
scratch extraction} and is applied the first time an extraction is done on a page.
It consists of three steps: (1) fragmentation,
(2) discarding rule application and (3) automatic pattern creation.
In the first step, fragments which contain clues
of the prices are found. In the next step, discarding rules previously designed
are applied to the fragments.
As a result, it is expected that only a fragment containing the
price remains.
From this fragment, in step three the price is extracted and a pattern
for future extractions is obtained.
The first step is inspired in the segmentation of~\cite{DBLP:dblp_journals/tkde/ZhaiL06}
while, to the best of our knowledge, there is no proposal of discarding rules in the
literature.

The second kind of extraction method is called {\em pointing pattern extraction} and
uses the pattern determined in the from scratch extraction for retrieving the price. 
This second method is applied the second and subsequent times an 
extraction is done on a page. If this extraction
does not find any price, the process begins again and a from scratch extraction is applied.

The difficulty of the problem we propose strives in the fact that in the detail web pages
there are usually many candidate values from which only one is a true positive and the
rest are true negatives. For example, in the page of {\em The Distant Hours} paperback book in Barnes \& Noble,
there is one true positive and fifty-six true negatives. Then, the question is:
how to find the right discarding rules and the correct order of application so that
only the true positive remains? For example, 
the finding of semantic discarding rules clashes with the problem of meaning. For deciding whether to eliminate a fragment, 
we would like to identify the concept included in the fragment and, if the concept is not appropriate, the fragment should be discarded. 
However, the algorithm cannot deal with the concept and only knows how to deal with expressions of the concept. 
The problem is that there are a great number of expressions of the same concept, so we should codify 
in the algorithm all the expressions of the concepts which, once found, lead to the algorithm discarding the fragment.
Maybe in some special case we are able to identify all the expressions of a concept, but in general we do not know in advance all these expressions.
Thus, consider we find an expression of the concept `discount' in
a fragment. As a consequence, the fragment must be discarded because we are interested in the concept `price'.
However, consider the following fragments:

{\tt\small  
(a) <span>19,95\&euro;</span>  instead of  <span>21\&euro;</span>

(b) <div>Save 1,05\euro</div>}

One way to identify the concept `discount' is the presence of
the word `Save'. So, fragment (b) is easily discarded but
in fragment (a) it is more difficult to identify the presence
of the discount concept.

%%When none of the previously found patterns work, the maintenance
%%we propose consists of generating new patterns under the assumption
%%that the structure of the old page has changed

The contribution of this paper is twofold:
\begin{enumerate}
\item{We provide a novel method for extracting the prices in web pages 
based on the use of discarding rules (see Section 4). An analysis of the results
of the method is done in Section 6.}
\item{As for the final user, the main feature of Wextractor is usability. (S)he only has to signal the pages of his(her) interest 
by means of a button of a Chrome extension and to check the extracted prices in a web app
(see a more complete description in Section 3). 
No technical knowledge or programming is required from the user.}
\end{enumerate}

The rest of the paper is organized as follows. 
Section 2 is devoted to the related work.
In Section 3 we describe the  problem and the solution for extracting prices from web pages.
In Section 4 we explain the discarding rules. In Section 5
we detail the algorithms. 
%%Section 5 is devoted to some extraction examples. 
In the last Sections, we deal with
the experimental validation and conclusions and future work.

\section{Related Work}

There are a lot of works about data extraction from HTML documents and surveys
about these works such as~\cite{DBLP:dblp_journals/tkde/ChangKGS06,DBLP:journals/ftdb/Sarawagi08,DBLP:journals/corr/abs-1207-0246},
so in this short section we can not do justice to all contributions.

We distinghish two kinds of works: commercial tools and research papers.
Commercial tools like Mozenda~\cite{Mozenda}, iMacros~\cite{iMacros}, Visual Web Ripper~\cite{visualwebripper}, Lixto~\cite{lixto}
offer wrapper generation
frameworks for recording user actions and extracting data later on based on
the recorded information. In our proposal, the user does not have to install
another tool for extracting data, because it is enough to install a Chrome
extension. 

With regard to research papers, the book~\cite{DBLP:books/sp/dcsa/Liu07}
distinguishes three kinds of extraction: manual, wrapper induction and automatic extraction.
In the first, the human programmer writes a program to extract target data. An
example is~\cite{DBLP:dblp_conf/iiwas/StarkaHN13} and its main drawback is that it does not adapt well to HTML structure changes.
In the wrapper induction approach, a set of extraction rules is extracted from previously labeled
pages. Some examples are~\cite{DBLP:dblp_conf/ijcai/KushmerickWD97,DBLP:dblp_journals/ml/Soderland99,DBLP:dblp_journals/dke/LaenderRS02}.
Finally, the automatic extraction is an unsupervised approach where patterns are automatically
found for data extraction. Some examples are~\cite{DBLP:dblp_conf/vldb/MeccaCM01,DBLP:dblp_conf/icde/ArasuG03}. 
Our paper adheres to this proposal
because it overcomes the two shortcomings of manual labeling effort and of costly wrapper maintenance
of the wrapper induction approach.

To the best of our knowledge, the extraction of prices without programming
tasks by the users is novel and there are not any other methods to compare with.
Also, we have not found papers with method similar to our paper.
The first phase is inspired in the segmentation of~\cite{DBLP:dblp_journals/tkde/ZhaiL06}
while, to the best of our knowledge, there is no proposal of discarding rules in the
literature.
We simplify the follow-up of the price in a detail page
of interest because no programming is required from the user and because
we concentrate the information in the user account.
The majority of papers try to extract all the information of a page.
See, for example,~\cite{DBLP:dblp_journals/vldb/FurcheGGSS13}.

\section{The problem and our  solution}

We state our problem as follows: 

Given a detail web page {\tt\small pg} of an entity {\tt\small e} with url {\tt\small u}, 
extract the price of {\tt\small e} of page {\tt\small pg}. 

An example of the problem is the  query (q1): Given the web page \\ {\tt\small http://www.wiggle.co.uk/asics-gel-nimbus-19-shoes/}
devoted to the Asics Nimbus 19 shoes, find its price.

Note that when we say detail page, we do not mean that the page contains no other information. 
In fact, for example, the Asics Nimbus 19 shoes contains twelve more prices of running-related objects apart from the price
of the shoes.

Let us explain our solution, the web data extraction system Wextractor. 
It combines a web app and our method for extracting prices.
Before beginning to work, the user has a user account
in the web app and 
a browser extension is installed to connect with the user account. 
As a result, the button `Follow this price' 
appears in the browser.
When the user arrives at a page of interest, for example, the page of Asics Nimbus 19 shoes
in Wiggle, (s)he clicks the button `Follow this price'. Then,
the HTML code and the url of the page are taken and sent to the user account.

\begin{figure}
\centerline{\includegraphics[width=60mm]{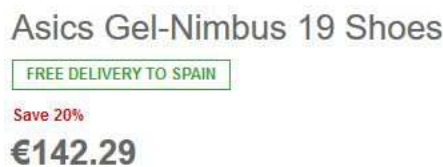}}
\caption{Fragment of Asics Nimbus 19 shoes web page as of May  2017}
\label{fig:interface1}
\end{figure}

Next, the from scratch extraction of our method starts.
The fragmentation step consists of detecting fragments of HTML code which contain clues indicating
the presence of prices.
Examples of clues are \&euro; or \$.
For query (q1), nine fragments are extracted, two of them being:

(f1) {\tt\small <div class="Wprice">\&euro;142.29</div>}

(f2) {\tt\small <div class="saving">SAVE20\%=\&euro;57.71</div>}

\begin{figure}
\centerline{\includegraphics[width=120mm]{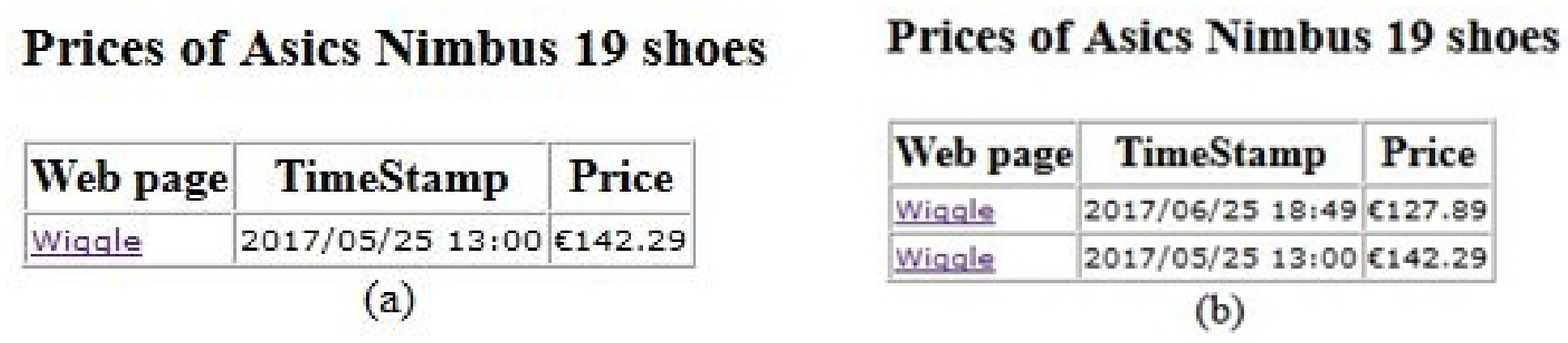}}
\caption{(a) View of the web app after the first extraction and (b) after the second extraction.}
\label{fig:interface2}
\end{figure}

In the second step, discarding rules are applied to the collected fragments where a discarding rule searches for properties
of the fragments which indicate that they do not contain the price of the entity {\tt\small e}.
An example of a discarding rule, called {\tt\small semr1}, is
`discard those fragments which contain the word SAVE'.
(f2) is an example of a fragment discarded by the {\tt\small semr1} discarding rule.
After applying the discarding rules, a small number of fragments remain.
From these fragments, in the third step, the target price is extracted.
In the example, the fragment (f1) is the only non discarded fragment and from it the
value 142.29 euros is obtained. This result is shown in the web app as in Figure~\ref{fig:interface2}(a).

If  the user wanted to check the price of an entity again, we would have two options. 
One of them is to repeat a from scratch extraction again. 
The second one is to try to use the information of the from scratch extraction for speeding up the process.
As our method relies on detecting HTML fragments and, in general, the web page structure 
does not change too much with time,  we have chosen the second option.

Following the second option, the non-discarded fragments of the from scratch extraction are used for building
a pointing pattern, that is, a regular expression that matches the price.
For example, after applying the from scratch extraction in query (q1), only the fragment (f1) remains. From this fragment,
the pointing pattern pp1 {\tt\small Wprice">\&euro;[0-9]\{2,3\}$\backslash$.[0-9]\{1,2\}} is built for subsequent extractions. 
It is worth noting that when a new extraction is done with this pattern, the price of 142.29 euros could have fallen below 100 euros. 
For this reason, the first part of the numeric pattern is {\tt\small [0-9]\{2,3\}}, that is, prices of less than 100 euros are also matched by the pattern.

%%%% next time
From the side of the user, the second and subsequent times (s)he wants to check the price of the entity, (s)he does not have to visit the page again . Instead of this,
 (s)he will enter into his/her account and (s)he will click the `Find again' button.
For this new extraction, 
the pointing pattern extraction is used for retrieving the price. For query (q1), the result
of the second extraction can be seen in Figure~\ref{fig:interface2}(b).

Having explained our general idea, technical details and algorithms
are offered in Sections 4 and 5.

\section{Discarding rules}

In this section, we present one of the main contributions of this paper, the discarding rules. 
Previously, we briefly define the notion of fragment.

A fragment is an HTML string which contains a price. The string is enclosed  between
an open and a closed HTML mark. The price is composed of a value and a clue for the
price as, for example, \$, EUR or \&euro;.

Basically, a discarding rule identifies a property which indicates that 
a fragment does not contain a price. Each discarding rule
is applied to all fragments and those fragments verifying the property are marked
as deleted, meaning that no right price is available inside it. For example,
a discarding rule identifies fragments beginning with the element strike, so that,
during the execution of the discarding rule, fragments with this property are discarded.

{\bf Definition}
A discarding rule is a Condition-Action rule which rules out  fragments 
as containers of a possible price.
Its syntax is:

{\tt\small  

discardingRule name (HTMLFragment f)

\hspace{.5cm}  IF condition  THEN action 

}

Some examples of discarding rules, with their conditions and their meaning 
are shown in Table~\ref{tab:features}. There are four types of discarding rules: 
syntactic, semantic, frequency and threshold. 
Let us explain each type of rule.

\begin{table}[t]%
\caption{Some examples of discarding rules}
\label{tab:features}
\begin{tabular}{|p{1cm}|p{5cm}|p{6cm}|}
\hline
Name&Condition & Meaning\\\hline
syr1&
beginsWith(`$<$strike', f) &
The fragment f begins with the string $<$strike \\\hline
syr2&
beginsWith(`$<$script',f)&
The fragment f begins with the string $<$script\\\hline
semr1&
containsNear(`Save',f)&
The fragment f contains the string `Save' near the price\\\hline
fr1&
nPre(f)$>$=3&
The number of fragments with the same pre as f is greater than three\\\hline

\end{tabular}
\end{table}%

\subsubsection{Syntactic rules}
The syntactic rules are related to the meaning of the HTML elements as well as to the meaning of the styles included in CSS files. 
Some syntactic rules and the rationale behind them are:

{\tt\small discardingRule syr1(HTMLFragment f)

\hspace{.5cm} IF beginsWith(`<strike', f) THEN markAsDeleted(f)}

Reason for the rule: The mark strike denotes that the price
is crossed-out, so it is no longer valid. 
It can mean, for example, an old price for the entity.

{\tt\small discardingRule syr2(HTMLFragment f)

\hspace{.5cm} IF beginsWith(`<script', f) THEN markAsDeleted(f)}

Reason of the rule: A precondition of our extraction is that the HTML page is not generated by Javascript, 
so, the content of the scripts is not visualized in the page.

\subsubsection{Semantic rules}

The semantic rules are related to the meaning of the words contained in the fragments.
An example of semantic rule  is:

{\tt\small discardingRule semr1(HTMLFragment f)

\hspace{.5cm} IF containsNear(`Save', f) THEN markAsDeleted(f)}

Reason of the rule: The quantity which appears in the fragment corresponds to
a discount and not to a price.

\subsubsection{Frequency rules}

These are established according to the repetition of specific strings in the fragments.
Some examples of discarding rules and the rationale behind them are:

Rule fr1. For each fragment f, let nPre(f) be the number of fragments whose pre
coincides with the pre of f, where the pre is the start tag including attributes.

{\tt\small discardingRule fr1(HTMLFragment f)

\hspace{.5cm} IF nPre(f)>=x THEN markAsDeleted(h), such that  pre(h)=pre(f)}

The value of x is obtained by direct observation of real cases.
A typical value is x=3.

Reason for the rule: The pre of the fragment which contains the 
price is very specific. So, if x or more fragments have the same pre, all of
them are discarded.

Rule fr2. For each fragment f, let nChar(f, n) be the number of fragments
whose first n characters coincide with the first n characters of f.

{\tt\small discardingRule fr2(HTMLFragment f)

\hspace{.5cm} IF nChar(f, n)>=x 

\hspace{.5cm} THEN markAsDeleted(h), such that  substr(h, 1, n)= substr(f, 1, n)}

The values of n and x are obtained by direct observation of real cases. Typical values
are n=21 and x=3. 

Reason for the rule: If there are several fragments with the same beginning, they may
be discarded because the beginning of the fragment containing the price
is usually very specific.

\subsubsection{Threshold rules}
These rules filters the prices based on limits imposed by the user or limits given by common knowledge. 
For example, if the entity is a new car, the common knowledge tells us that its price is greater than 6000 euros. 
Moreover, the user can tell the maximum price (s)he can allow is 12000 euros. 
So, all cars out of the range 6000-12000 euros must be discarded.

{\tt\small discardingRule thresr1(HTMLFragment f, Number min, Number max)

\hspace{.5cm} IF (value(f)<min or value(f)>max) THEN markAsDeleted(f)}

\section{The Algorithms}

In this section, we present the algorithms which implement our proposal. The from scratch extraction,
that is, the first time we search for a price, is implemented by means of the
{\tt\small doFromScratchExtraction} algorithm (see Section~\ref{fromscratchalgorithm}).
When pointing patterns are available, we do a pointing pattern extraction
implemented by the algorithm {\tt\small doPointingPatternExtraction} (see Section~\ref{pointingalgorithm}).
The main algorithm (see Section~\ref{main}) decides which of the previous
algorithms is executed.

\subsection{The doFromScratchExtraction algorithm}
\label{fromscratchalgorithm}
In this extraction, first the clues are detected in the web page.
Examples of clues are \&euro; ~EUR  ~\euro ~\&\#8364;.
In the following steps, fragments containing the clues 
are detected and discarding rules are applied to these fragments.
As a result of applying the discarding rules, there remains a collection of fragments where
prices appear. These values are considered possible prices. 
If  n is the total number of distinct found values, these are the results of the algorithm:

n=0. There is no price on page pg.

n=1. The price in page pg is the only value found.

n$>$1. There are several prices. 

In some pages, such as the Asics Nimbus 17 shoes page of Amazon, 
there can be two values indicating the minimum and maximum price.
Our algorithm detects this situation and considers that there is only one  value, given
by the pair (minimum, maximum).

The complete algorithm can be seen in Table~\ref{tab:fromscratch}.

\begin{table}[t]
\caption{Algorithm for doing a from scratch extraction} \label{tab:fromscratch}
\begin{tabular}{p{0.25cm}p{12cm}}
& Algorithm doFromScratchExtraction\\\hline
& Input: HTML code of page pg 
\\

& Output: 

 list of candidate prices

 pointing pattern, if any

Preconditions:

  The page pg is available

  The HTML code is explicitly available. It is not generated, for example, by means of Javascript
\\\hline

& Pseudocode\\

  1.  & {\tt\small  \hspace{3mm}listOfClues<--loadClues()}\\

  2.  & {\tt\small  \hspace{3mm}listOfDiscardingRules<--loadDiscardingRules()}\\

  3. & {\tt\small  \hspace{3mm}for each clue c of listOfClues do}\\

  4. & {\tt\small  \hspace{4.5mm}fragments<--findAssociatedFragments(HTMLCode, c)}\\

  5. & {\tt\small  \hspace{4.5mm}listOfFragments<--add(listOfFragments, fragments)}\\

  6. &{\tt\small  \hspace{3mm}endfor}\\

  7. &  {\tt\small  \hspace{3mm}for each discardingRule r of listOfDiscardingRules do}\\

  8.  &   {\tt\small  \hspace{4.5mm}applyDiscardingRule(r, listOfFragments)}\\

  9.  & {\tt\small  \hspace{3mm} endfor}\\

  10.  &{\tt\small  \hspace{3mm}listOfCandidateFragments<--getCandidateFragments(listOfFragments)}\\

  11.  & {\tt\small  \hspace{3mm}listOfValues<--extractValues(listOfCandidatesFragments)}\\

  12. & {\tt\small  \hspace{3mm}if (\#listOfValues=1) then}\\

  13. & {\tt\small  \hspace{4.5mm}pp<--extractPointingPattern(listOfCandidateFragments)}\\

  14. & {\tt\small  \hspace{3mm}end if}\\
  
  15.  & {\tt\small  \hspace{3mm} return (pp, listOfValues)}\\

\end{tabular}

\end{table}

\subsubsection{Algorithm description}
In line 1, a list of clues previously prepared by the programmer is retrieved 
and it consists of strings as {\tt\small ~EUR  ~\euro ~\&\#8364;} and so on. 
In line 2, the discarding rules are loaded.
In lines 3 to 6 for each occurrence of each clue in the HTML code, 
associated fragments to the instances of the clues are found. Next, discarding rules 
are applied to the fragments (lines 7 to 9).
Those fragments which have not been discarded by the application of the rules,
make up the list of candidate fragments (line 10).
From this list of fragments, the list of prices is extracted (line 11).
In lines 12 to 14, a pointing pattern to be used for subsequent extractions is determined.
The pointing pattern and the remaining prices are returned as results in line 15.

%%{\bf Examples} See Section~\ref{examples}

\subsection{The doPointingPatternExtraction algorithm}
\label{pointingalgorithm}
When a from scratch extraction is done, it is sure that the web page is available because the search is executed just after the user has selected the page. 
However, when a new search  is done on the same page by using the pointing pattern extraction, it can happen that the url no longer points
to the page or that the structure of the HTML code 
of the page has changed so that the pointing pattern does not find any result. We impose as a precondition for this algorithm
that the page pg is available. The sketch of the subalgorithm
can be seen in Table~\ref{tab:find}.

%%Había escrito esto al margen, pero no sé dónde encajarlo aquí:
%% If the page is not available, nothing can be done, neither with the from scratch search nor with the pointing pattern search.

\begin{table}[t]
\caption{Algorithm for doing an extraction based on a pattern} \label{tab:find}
\begin{tabular}{p{0.25cm}p{12cm}}
& Algorithm doPointingPatternExtraction\\\hline
& Input: HTML code of page pg,  
pp pointing pattern with which we search the price in page pg\\
& Output:  list of prices\\

& Precondition: The page pg is available
\\\hline
& Pseudocode\\

1.&  {\tt\small  \hspace{3mm}lstrings<-match(HTMLCode, pp)}\\

2. &  {\tt\small  \hspace{3mm}for each x in lstrings do}\\

3. &   {\tt\small  \hspace{4.5mm} v<--extractValue(x, pp)}\\

4. &    {\tt\small  \hspace{4.5mm}lvalues<-add(lvalues,v)}\\

5. &  {\tt\small  \hspace{3mm}end loop}\\

6. &    {\tt\small  \hspace{4.5mm}return lvalues}\\

\end{tabular}

\end{table}

\subsubsection{Algorithm description}
In line 1, the strings matched by the pointing pattern pp are retrieved.
In lines 2 to 5, for each matched string, the price
found in the string is extracted and accumulated into variable lvalues.
In line 6, this variable is returned as result of the algorithm.

%%{\bf Examples} See Section~\ref{examples}

\subsection{The findAttributeValues algorithm}
\label{main}
This is the main algorithm and
combines the two algorithms we have described in the previous sections. 
First, a pointing pattern extraction is done (lines 1 to 11). 
If the extraction does not give the expected results, 
then a from scratch extraction is done (lines 12 to 19). 
In particular, the first time an extraction is done, as there are  no previously available patterns, 
a from scratch extraction is always done.
The sketch of the subalgorithm
can be seen on Table~\ref{tab:main}.

\begin{table}[t]
\caption{Algorithm for finding the price of a detail page} 
\label{tab:main}
\begin{tabular}{p{0.25cm}p{12cm}}
& Algorithm findAttributeValues\\\hline
& Input: extraction kit sk, composed of the url of the page and of pointing patterns\\
& Output: 

(false, -1) the page pg is not available

(false, -2) there are many prices found in page pg

(false, 0) there are none price found in page pg

(true, v)   v is the unique price found in page pg
\\\hline
& Pseudocode\\

1.&{\tt\small  \hspace{3mm} if existsPage(sk.u) then}\\

2.&{\tt\small  \hspace{4.5mm} HTMLcode<--findHTMLCode(sk.u)}\\

3. &{\tt\small  \hspace{4.5mm} s <- empty}\\

4. &{\tt\small  \hspace{4.5mm}pp<-- lastpp(sk)}\\

5. &{\tt\small  \hspace{4.5mm} while s is empty AND pp is not empty}\\

6. &{\tt\small  \hspace{6mm} s<--doPointingPatternExtraction(HTMLCode, pp)}\\

7. &{\tt\small  \hspace{6mm}pp<--nextpp(sk, pp)}\\

8. &   {\tt\small  \hspace{4.5mm}end while}\\

9. & {\tt\small  \hspace{4.5mm}if \#s=1 then  return (true, s[1])}\\

10. &   {\tt\small  \hspace{4.5mm}elsif \#s>1 then return (false, -2)}\\

11.&   {\tt\small  \hspace{4.5mm}else}\\

12.&     {\tt\small  \hspace{6mm}(pp, s)<--doFromScratchExtraction(HTMLCode)}\\

13.&       {\tt\small  \hspace{6mm}if \#s=0 then return (false, 0)}\\

14.&       {\tt\small  \hspace{6mm}elsif \#s>1 then return(false, -2)}\\

15.&      {\tt\small  \hspace{6mm}else}\\

16.&       {\tt\small  \hspace{7.5mm}addPointingPatternToExtractionKit(pp, timestamp, sk)}\\

17.&       {\tt\small  \hspace{7.5mm}return (true, s[1])}\\

18.&     {\tt\small  \hspace{6mm}end if}\\

19.&   {\tt\small  \hspace{4.5mm}end if}\\

20.& {\tt\small  \hspace{3mm}else}\\

21.&   {\tt\small  \hspace{4.5mm}return (false, -1)}\\

22.& {\tt\small  \hspace{3mm}end if}\\

\end{tabular}

\end{table}

\subsubsection{Algorithm description}
In line 1, the existence of the page pg is checked. 
If the page is not available, the value (false, -1) is returned (line 21). 
Otherwise, in line 2, the code of the page is retrieved.
For each pointing pattern of the extraction kit, an extraction is done (lines 4 to 8), begining with
the latest pointing pattern, by means of
the algorithm {\tt\small doPointingPatternExtraction} (see Section~\ref{pointingalgorithm}).
When an extraction finds a price, the extraction finishes.
At the end, the variable s stores the prices found for the first pointing pattern
which has found prices.
Then, if s has only one element, this element is returned 
as result(line 9). If s has more than one element, the value
(false, -2) is returned, indicating that there are many prices. 
If s has no elements, this means
that the pointing pattern extraction did not find any solution.
This can be for many reasons as, for example, (1) it is
the first time the extraction is done and there are no patterns,
or (2) there can be patterns but the structure of the page
has changed so that no old pattern finds the correct price.
In any case, the existing patterns are not deleted because they can
be used again later. As the extraction did not find any solution, 
then a from scratch extraction is done (line 12).

As a result of this extraction, no price is found and the value (false, 0) is returned (line 13) or more than
one price is found and the value (false, -2) is returned (line 14)
or exactly one price is returned (line 17).
Before returning this value, a pointing pattern is determined (line 12) and will be used in subsequent pattern based extractions. 
For this purpose, the pattern and its timestamp are added to the extraction kit (line 16).

\section{Experimental validation}
\label{experimental}
In this section we present the empirical evaluation of the Wextractor method.
To evaluate the effectiveness of our solution, we searched for (1) price trackers and
(2) adequate datasets on the web.
With respect to the former, we found price trackers in the web with the same goal as Wextractor. 
However, all the trackers we have visited (slickdeals, pricegrabber,
camelcamelcamel) are able to track a limited number of sites, unlike Wextractor,
which can track prices in any site. So, we searched for adequate datasets on the web.
After an extensive search, we only found the dataset of paper~\cite{hao2011one},
composed of around 124K pages collected from 80 websites. The websites
are related to 8 vertical, including autos, cameras or books.
For each vertical, 10 popular websites were identified by issuing queries
to search engines. For each website, 200-2000 detail web pages were downloaded.

This dataset is also insuffcient because only the camera and the auto verticals include
the attribute price. So, we decided to create our own dataset. 
For this purpose, we gathered sites the page myalerts.com (now trackif.com) claimed
to be able to track. They are the most active e-commerce sites on the web. For these sites,
we selected those which accomplished the following properties: (1) were accesible
on the web, (2) offered detail pages and (3) the prices of the objects in the detail pages
were available in the HTML code (not in script code). Finally, we have worked with 444 sites.
The complete dataset can be downloaded at~\cite{dataset}.

For each of these sites, we selected one detail page and we annotated, by direct inspection,
the price of the corresponding entity. 
%%Our complete dataset together with exhaustive information 
%%about the detail pages, the prices and the results of the experiment can be seen at+++.
Then, we executed Wextractor for each page and we compared the results. 
We evaluated our method against our dataset by using precision and specificity.
As there are many true negatives and only one true positive, we have chosen the specificity to
measure the true negative rate. In general, the experimental results show that our algorithm discovers
prices in different web sites. In average, it achieves 80.00\% precision and 97.00\% specificity. 
It returns perfect results (100\% precision and specificity) for 342  out of 444 websites of the dataset.
It returns 100\% precision for 342  out of 444(77\%) of the  websites of the dataset.
It returns 100\% specificity for 387  out of 444 (87\%) of the   websites of the dataset.

We have implemented Wextractor by means of the Oracle database 11g and the PL/SQL
programming language. Packages like UTL\_HTTP have been of great help in this task.

\section{Conclusions and Future work}

In this paper we have presented a web extractor system, 
called Wextractor, whose purpose is doing a follow-up
of prices in web pages.
Several questions arise from this work and will be dealt with in
future papers:
Could our three steps method be extended to other problems
as defined in~\cite{DBLP:journals/ftdb/Sarawagi08}, namely,
entity extraction or binary relationship extraction?
Could our proposal be generalized to other attributes different from the attribute price?

\bibliography{websearch}{}
\bibliographystyle{plain}

\end{document}